\documentclass[12pt,a4paper]{article}

\usepackage{graphicx}\usepackage{amsmath}

\usepackage{subfigure}

\begin{document}

\begin{center}{\bf \Large Some recent attempts to simulate the Heider balance problem}\\[5mm]

{\large  Krzysztof Ku{\l}akowski \\[3mm]

\em {Faculty of Physics and Applied Computer Science, AGH University of Science and Technology, al. Mickiewicza 30, 30-059 Krak\'ow, Poland\\

E-mail: kulakowski@novell.ftj.agh.edu.pl\\\today}}

\end{center}

\begin{abstract}

The Heider balance is a sociological problem of a division of a community into hostile groups, where all interpersonal relations within the groups are friendly and all relations between the members of different groups are hostile.  Here we trace how the research of the process of attaining the Heider balance has moved during last ten years from a psycho-social to a computational problem.

\end{abstract}


\section{Introduction}

Who is enemy and who is friend? The skill to distinguish them was one of most basic in the evolution of the human race, and still it is helpful. If the surrounding world can be divided into two compact sets of friends and of enemies, the situation is clear. However, often our friend is also a friend of our enemy; can we trust him? Then we suffer from a cognitive dissonance, what makes our life harder. To regain comfort, we are able to sacrifice a friendship or sometimes a common sense; we are also able to believe to a politician just because he claims he also believes in our worths. Then, the Heider balance can concern people, ideas and media.

The problem has been formulated by Fritz Heider (born 111 years ago) in 1944 \cite{hei1,hei2} in terms of local triads. Assigning a positive or negative (friendly or hostile) sign to a reciprocated relation between each two of the triad members, we get three signs. Basically, there are four possibilities: $(+,+,+)$, $(+,+,-)$, $(-,+,-)$ and $(-,-,-)$; for the sake of this discussion we have no reason to distinguish $(+,+,-)$ and $(+,-,+)$ etc. Two configurations are balanced: in $(+,+,+)$ everybody likes everybody and there is only one group, in $(-,+,-)$ one person is hostile to the remaining pair of friends. Two other configurations are unbalanced: in $(+,+,-)$ there is one hostile relation between two triad members, still connected by friendship with the third person. In $(-,-,-)$ all relations are hostile, and it is somewhat strange for every triad member why the remaining two - both abominable - dislike each other. Then, also the configuration $(-,-,-)$ is considered as unbalanced.

As it was recognized by mathematicians \cite{har}, the attaining of Heider balance in local triads brings heavy consequences. Provided that relations exist between each pair of members, all relations between members of the same group are positive, and all relations between members of different groups are negative. In other words, the balance appears not only in local triads, but globally. In this way, the removal the unbalanced relations of individual persons happens to be irrevocably dysfunctional for the community. The equivalence of the local and the global balance is the content of what is known as the structure theorem \cite{har}. It was generalized in Ref. \cite{dav} to include the $(-,-,-)$ configuration as a balanced one; in this case more than two hostile groups can appear.

Although the stationary state is mathematically determined, it is far from obvious how the social system evolves towards the balanced state. Within the social psychology, most natural approach is the participating observation; after some months, a report can be written with careful analysis of sequence of events, supported by personal characteristics of the group members. In mathematical modelling, the group members are transformed into nodes of a graph, and the interpersonal relations into links. Obviously, the sociophysicists with their computer simulations and phase transitions spontaneously aggregate into mathematical sociology, stretching in this way the definition of social sciences over its former boundaries. 

The aim of this text is to facilitate this invasion even further. To achieve this, simple computational motifs are selected from some recent works on the Heider balance. These motifs are listed below as subsequent sections. Some questions seem to be shared by more than one approach: Is the balance always attained? What is the nature of its attaining? How to slow it down or stop? What is the role of the group size? Is the final state unique? If not, what it depends on?

\section{More than two clusters}

The problem set in Ref. \cite{dor} is to look for a partition of the network into clusters as to minimize the total number of negative links within and positive links between clusters in the given distribution of positive and negative signs of the links. The input data is a network of $N$ nodes and some directed links $R$ between them. The procedure is a random walk in the state space of the partitions. During the search, the number $k$ of clusters remains fixed. The flowchart is as follows:

1. For given value of $k$ select randomly a partition $C$ of the graph nodes into $k$ clusters. 

2. Find in the original network $R$ the number $N_n$ of negative links within the clusters of $C$ and the number $N_p$ of positive links between the clusters of $C$. The work function is $W_R(C)=N_n+N_p$.

3. Select randomly a neighbour $C'$ of the partition $C$, obtained either by moving a node from one subset to another, or by interchanging two nodes from different subsets. Find the work function $W_R(C')$.

4. If $W(C')<W(C)$, accept $C'$ as a current partition $C$. 

5. Repeat steps 3 and 4 many times.

As reported in Ref. \cite{dor}, an example of $N=18$ nodes (see below) required a dozen of runs of the algorithm, each of 5500-6800 repetitions of steps 3 and 4. Applying the algorithm, one can obtain a set of partitions $C(k)$ which give minimal values of the work function. The output value of $k$ is where the work function gets its minimal value. However, as pointed by the authors of Ref. \cite{dor}, it is possible that the partitions found are not optimal. As we know both from biological and magnetic considerations \cite{kauf,stauf}, the step 4 of the above algorithm leads to a local equilibrium in the space of partitions, where the work function $W(C)$ is not minimal. 

As an example, the authors discuss the Sampson monastery data \cite{sam} on the relationships between 18 young monks. These relationships were measured by Sampson at consecutive times $T_2$, $T_3$ and $T_4$. (The measurement at $T_1$ was done before a new group of monks appeared and therefore these results were not relevant for further analysis). Each time, the monks were asked to indicate names of three others which they liked most and three which they disliked most. At each time $T_i$, the relations were written into a directed relation matrix. The matrix elements $r_{ij}$ were $-3,-2,-1$ and $1,2,3$ to indicate six monks selected by an $i$-th monk, from those most disliked to most liked, respectively. Otherwise, $r_{ij}$ =0. These data are conversed to $\pm 1$.  The procedure given above, when applied to the matrices, gave the results as follows: each time, the number of clusters $k=3$ gave minimal values of the work function. Further, the partition obtained at $T_2$ as unique was consistent with the psychological observation done by Sampson. Further, this partition was reproduced when analysing the data at $T_3$ and $T_4$. The value of the work function calculated for these optimal solution(s) decreased in time almost twice.

The calculation was repeated when taking into account the value of the links, not only their signs. This modification was introduced via weights of the links in the work function. The only difference in the results was that the data for $T_4$ gave two partitions; one the same as before and one slightly modified. As indicated by the authors of Ref. \cite{dor}, in general valued or signed links make a difference; however, in the investigated case the results appeared to be particularly robust.

\section{Role of individuals reconsidered}

More recently, an attempt was made to include the actions of individual network members to the process \cite{hum}.  The method of the calculations is discussed in two stages. At first, the authors describe the multi-thread aspect of the model. The algorithm includes the decisions of the network members: each agent tends to change the links directed from its node to its neigbours in such a way as to reduce imbalance. To do this, one can modify the link state, which can be negative, zero (no link) or positive. These individual decisions are supported by a global process where a balanced network is found as close as possible to the current configuration \cite{dor}. This information is sent to all agents and serves as a guide in the case of ambiguities; still, as indicated many times by the authors of Ref. \cite{hum}, the final state obtained by the algorithm varies from one realization to another, even if the initial state of the networks was the same.

Although in the description of their method the authors clearly differ between the multiple thread model and the discrete event simulation model, the algorithm applied joins features of both. The idea of the discrete event simulation model is that instead of working in real time, it modifies the states of the links sequentially one by one. The sequence order is determined by the algorithm as well; once a future modification appears, it is inserted into the priority queue. Once the structure is balanced both for individuals and for the whole network, the calculation is stopped.

The main calculations of Ref. \cite{hum} are described in Section 4. The parameters of the calculations are: the group size $n$ (from 3 to 10), the initial probability $p$ of a negative link in the network, and the communication method. The list of the possible methods includes four items: {\it i)} dyadic method, where only a neighbour joined by the modified link is informed, {\it ii)} tell-friends method, to inform only neighbours joined with the informer by positive ties, {\it iii)} tell-acquaintances method, to inform all neigbours, {\it iv)} broadcast communication, to inform all the network. These methods provide different amounts of information for the agents about the actual state of the network; therefore we get the differences in the percentages of agents which perceive the network as balanced (if they are properly informed or not, depends on the actual method of communication). Other measured quantities are: the number of decisions of the agents needed to reach a stationary state, the level of the group imbalance, measured by a number of links necessary to be changed to get a balanced state, and the number of clusters in the final partition.

Some of the results can be seen as obvious, as for example the number of steps to get equilibrium increases with the system size. On the other hand, the same quantity shows a maximum with the percentage of links which are initially negative; this result has no clear explanation. Also, the level of imbalance decreases with the same percentage, what seems to be a counterintuitive result. In the discussion, the authors refer to several sociological threads, which could not be reflected in their simulation. In particular they indicate, that in a society the process of attaining balance is usually preceded by a development of reciprocity and transitivity of human relations \cite{dkks}. Although the latter reference deals also with the Heider mechanism, the simulation does not play an important role in that text; therefore we do not discuss it here. We only mention that the text reviews yet another set of sociometric data (pseudo-fraternity data) of Newcomb \cite{new,dr2}.

In summary, the authors state that any limitation of the simulation to individual decisions of the group members, although possible, "denies the social aspect of the balance processes". Instead, they call for a a model, in which "both social choice mechanisms and the group partitioning process are part of an actor's cognitive processing." More extensive discussion of sociological principles of modelling networks can be found in \cite{dr2}. However, this discussion deals with the conceptual formulation of the theory rather than with its mathematical or algorithmic formulation. Anyway, it seems that the text \cite{hum} triggered off research on the dynamics of the process.

\section{Monte Carlo dynamics}

Despite the above mentioned warnings, a numerical proof that the balancing process can be guided exclusively by individual decisions appeared in the same year 2003 \cite{zth}. The authors investigated the dynamics of attaining the balance state. They explained their motivation as follows: It is clear from the structure theorem, that a balanced system is in an equilibrium, but how we know that it ever attains this state? To deal with this dynamics, two simulations are reported in Ref. \cite{zth}. 

For the first one, the starting configuration of the network is represented by a random symmetric matrix of zeros and $\pm 1$ as the matrix elements. These elements represent no relation, positive or negative relation respectively. Then, one triad after another is inspected if it is balanced or not. If the product of three bond signs is negative, the sign of one bond, randomly selected from the three, is changed. If a triad has one null relation, it is corrected as to create a balanced triad. If there are more null relations, the triad remains unchanged. In this way the authors inspected some thousands of networks of 9, 16 and 25 nodes, each with 300 iterations. As the result, they got that all investigated samples evolved to the balanced state where at most two subgroups emerged. All bonds within each subgroup were positive, and all bonds between subgroups were negative. Next, the size of the larger group was usually larger when most bonds were positive in the initial state. When most bonds were negative in the initial state, the sizes of two obtained groups were approximately equal. According to the authors, this particular results could be a mathematical artifact. The next result was that a relatively small amount of negative relations in the initial state produced a splitting. In general, the sizes of the emerging subroups varied from sample to sample even if the initial proportions of negative to positive to null bonds were the same.

To investigate this particular result more thoroughly, the second simulation was designed. By controlling the set of subsequent pseudorandom numbers, the authors were able to investigate an influence of the initial state and of the order of selecting triads to repair on the final state. The simulation was performed for two cases: {\it i)} in the presence of null links in the initial state, and {\it ii)} for the fully connected initial state. In the case {\it i)} both initial configuration and the order of triads were found to be relevant. In the case {\it ii)}, however, the order of the triads to balance was found to influence neither the size of the obtained subgroups, nor even their content. In other words, the same nodes appeared in the same subgroups in the final state, whatever was the order of investigated triads.

This result is very particular, as one easily can find an example when this order does matter. Not to deal with a single triad, let us consider the case of four nodes, each connected with each. If initially all links are negative, all triads are to be balanced. The sequence of decisions can be, for example, as follows: triad 124, link 12 changed to be positive; triad 234, link 23 changed to be positive; triad 134, link 13 changed to be positive. We obtained a balanced state 123 vs 4. Simple permutation of nodes (234, 34 changed; 231, 31 changed; 214, 14 changed) gives a balanced state 134 vs 2; and so on. We have to conclude, that for some reasons no such example appeared in the reported simulations, although each initial state was checked with 300 different orderings.

Despite this question mark, the formulation of the problem given in Ref. \cite{zth} opened a way for continuators. We will return to it in Sections 6 and 7.

\section{Detecting communities}

The Heider balance can also be seen as a member of a broad class of problems of detecting communities. This subject is much too wide to be reviewed here. However, at least Refs. \cite{fre,gir} should not be omitted, as they deal with two sets of experimental data relevant to our subject. 

Ref. \cite{fre} reports a thorough analysis of the attendance register on 14 informal social meetings of 18 women in Natchez, Mississippi, in 1930's. In principle, the correlation between the attendance of particular women allows to separate out two cliques. However, there are several possible methods of this analysis, and the results remain ambiguous. Twenty one different methods were applied to these data, and the results were compared to each other. Obviously, there is no argument that any clique appeared there, and the ladies could be surprised or even angry if they knew how their behaviour is interpreted. Then, the only criterion to evaluate this or that method is the criterion of mutual accordance of the results. In this way, a method is supposed to be the best if its result agrees with the largest number of results of the other methods. In this competition, direct observation performed in 1930's by five ethnographers appeared to be somewhat less efficient. Six other methods got the highest rank. An earliest (1972) systematic method was just to minimize uncertainty due to departures from some prescribed division. This approach gave an optimal division. Other equally succesful methods (1991-2001) relied mostly on algebraic eigendecomposition or on an application of the genetic algorithm in the space of divisions, to maximize e.g. correlation between some ideal division and the observed data. 

Some of these methods allowed to reproduce not only the division, but also the leaders of two "opponent" groups. This evaluation was also the subject of a more recent work \cite{bol}, which refers to the network model of the community. Once the system is driven to attain the Heider balance, the centrality index is evaluated for each node as the appropriate component of the eigenvector of the connectivity matrix to the largest eigenvalue. If the sign is changed of the connectivity matrix elements between two blocks and of the eigenvector components for one of the block, the eigenequation remains true. Then, the value of the positive (negative) component of the eigenvector indicates the position of a given node in hierarchy in first (second) block or group. The connection of the centrality index to the position in the social hierarchy is not new; Ref. \cite{bol} shows how this connection looks like in the state of the Heider balance.

Other index of position of a node in the network is the betweenness - the level of contributing in the transport of an information between other nodes, hence the term: "how one is between". In Ref. \cite{gir}, three definitions of the betweenness are proposed. First definition (shortest-path betweenness) is based on the calculation of the shortest paths between each pair of nodes in the network. For each link, the betweenness is calculated as the number of shortest paths which run along the link. Second index (random-walk betweenness) counts how many random walkers pass through a given link. To evaluate the third kind (current-flow betweenness) one has to find the current along each link, when the source and sink are placed sequentially at each pair of nodes. Assuming that some kind of sink is present also in the case of random walk, Newman proved that the second and the third definition are equivalent \cite{gir,nwm}. After evaluating the index calculated with one of these methods, the link with the largest betweenness is cut. Then the betweenness of all remaining links are calculated again, and so on - until two groups appear. 

This algorithm was applied in Ref. \cite{gir} to the example known as the Zachary karate club. As described in \cite{gir}, Zachary investigated social relations between 34 members of a karate club at an American university \cite{zcr}. Zachary was able to write down these relations in the form of the $34\times 34$ connectivity matrix, indicating who contacts with whom. During this research, a conflict appeared between the administrator and the teacher, and the club happened to divide into two groups. This division, i.e. the content of both groups, was reproduced by the calculation of the authors of Ref. \cite{gir} exactly when the random-walk betweenness was used. The calculation of the shortest-path index misclassified only one node of the graph.

\section{Continuous dynamics}

As long as the relations between nodes are described with discrete variables, the results do depend on the order in which the unbalanced triads are repaired. A simple example of this dependence was demonstrated in Section 4. This ambiguity is removed if we use a set of differential equations, one equation for each link; then all relations evolve simultaneously. Such a set was proposed in \cite{pg1}, namely $dx_{ij}/dt=g_{ij}\sum_kx_{ik}x_{kj}$, where $x_{ij}=x_{ji}$ describes a symmetric relation between $i$ and $j$, and the factor $g_{ij}=1-(x_{ij}/R)^2$ assures that $x_{ij}$ remains finite. In a fully connected network, the sum is performed over all nodes $k$ different than $i$ and $j$. Initial values of $x_{ij}$ are drawn randomly from the range $(x_m-\epsilon,x_m+\epsilon)$. As a default, $x_m$ was set to be zero. The departure from the balanced state was measured as the number $N_g$ of unbalanced triads. The calculations were performed for networks up to $N=500$ nodes.

As the result of the simulation, the state of the Heider balance was achieved in all investigated cases. Typically for $N>100$ the curve $N_g(t)$, initially flat, decreased abruptly to zero at some time $\tau \propto N^{-1/2}$. Near to this time moment, the time dependence of $x_{ij}$ also changed from slow and seemingly random changes within the initial range $(-\epsilon, +\epsilon)$ to almost immediate jumps to the limit values $\pm R$. We remind that the definition of the balanced state takes into account only the signs of $x_{ij}$, and not their absolute values. Then the result on the final polarization of opinions to $\pm R$ could be treated as a supplementary information. In Ref. \cite{pg1}, this polarization was compared to variations of public opinion in Poland on the vetting law in 1999.

Within the same approach \cite{pg2}, the phase transition was found at $x_m=0.036$ from the state of two final clusters to one cluster with all $x$'s positive. Other parameters were: $N$=100, $\epsilon$=0.5, $R$=5.0. The time of getting equilibrium was found to have a maximum at $x_m=0$. This result was analogous to the finding of \cite{hum}. Finally, the calculation scheme was applied in Ref. \cite{pg3} to two examples described above: the case of women of Natchez \cite{fre} and the case of the Zachary karate club \cite{zcr}. In the first case, the calculation was done with the initial values of $x_{ij}$ equal to the correlations matrix calculated for the attendance of the women on their meetings. The final division was the same as obtained by the six best methods, reported by Freeman in Ref. \cite{fre}. In the second case, the result reproduced exactly the 'experimental' partition, as observed by Zachary \cite{zcr}.

Recently, an attempt was made \cite{pg4} to generalize the continuous description by an inclusion of asymmetric relations, {\it i.e.} for the case when $x_{ij}\ne x_{ji}$. In some cases, long-living pseudoperiodic behaviour of $x_{ij}(t)$ was found, with the probability which increased with the number of nodes. During this kind of the system evolution, the number of the unbalanced triads remains positive. Although the numerical simulation does not allow to state that in a given case the Heider balance will never be attained, the observed times to getting this balance are sometimes surprisingly long. The histogram of this times seems to display a fat-tail behaviour at least for $N$=3, where a relatively good statistics could be obtained.

\section{Stable unbalanced states}

The dynamics of getting the balanced state was investigated also very recently in Refs. \cite{red1,red2}. Numerical simulation was only a part of this extensive study, largely devoted to the constructed rate equations. As the present report is devoted to simulations, here we extract only a small part of this work. The algoritm differs between triads $\Delta_j$ with $j$ negative links, then triads $\Delta_0$ and $\Delta_2$ are balanced, $\Delta_1$ and $\Delta_3$ are not balanced. If a randomly selected triad is $\Delta_3$, one of its edges is changed and we get $\Delta_2$; if it is $\Delta_1$, it is transformed into $\Delta_0$ with probability $p$ or into $\Delta_2$ with probability $1-p$. Alternatively, another version of the algorithm is used with the additional condition that the overall number of unbalanced triads cannot increase during the simulation. In this case, some "jammed" stationary states are observed where the balance is not attained. These states can be compared to local minima of energy in a complex energy landscape; the role of energy is played by the number of unbalanced triads.

The authors obtained several numerical results which agree with their theoretical considerations. In particular, the time to obtain the balanced state depends on the network size in different way for $p<0.5$, $p=0.5$ and $p>0.5$. The probability of the jammed state is found to decrease with the system size. A phase transition is observed from two balanced clusters to one in the final state, for the initial probability of positive links $\rho_0 \approx 0.65$. Theoretical prediction is that the transition should appear at $\rho_0 = 0.5$. For $\rho_0 = 0$, the two obtained clusters are more or less equal. This result reminds the findings of Ref. \cite{zth}, and therefore does not depend on a particular algorithm.

In conclusions, the authors refer to important historical coalitions in history of Europe in 1872-1907. Another example is the relationships among major cultures (African, Hindu, Islamic, Japanese, Latin America, Sinic and Western), inspired by the fameous Huntington theory on the clash of civilizations. Also, the authors indicate ways of further research in the field: indifferent interactions, continuous-valued interactions, asymmetric interactions, acceptance of triads $\Delta_3$ as balanced. At least some of these points are already met by the followers in Refs. \cite{rvyo,rvo}, where the analogy is exploited between the unbalanced networks and the frustrated spin glasses.

\section{Spin glasses and WWII}

In 1993 Axelrod and Bennett have formulated a model approach to analyse possible coalitions in a given set of countries \cite{axbe}. Although the paper does not refer directly to the Heider's works, the story is so close to our subject that it makes sense to remark it here.

The fixed parameters of the system are the size $s_i$ of each country counted by index $i$ and its propensities $p_{ij}$ to be close to other countries $j$. The search is to find an optimal division $X$ into two coalitions as to minimize the energy $E$ defined as $E(X)=\sum_{i,j}s_is_jp_{ij}d_{ij}(X)$, where $d_{ij}(X)=1$ if countries $i,j$ belong to different coalitions, $d_{ij}(X)=0$ otherwise. Analysing the sizes and the propensities of seventees European countries in 1936, the authors of Ref. \cite{axbe} were able to reproduce almost exactly the coalitions in the Second World War (the only wrong assignments were Portugal and Poland). The text \cite{axbe} operates with the term 'frustration', designed to mean the inability of a country to act according with its propensities.

This entrance to the area of statistical physics found a response \cite{gal}. The arguments used in the discussion can be found in (Brit. J. Polit. Sci. {\bf 28} (1998) p. 411, 413) and here we have nothing new to add. However, for our purposes one detail should be emphasized. As it was proven in Ref. \cite{gal}, the energy $E(X)$ can be translated into the Ising Hamiltonian at zero temperature, where the propensities play the role of positive and negative bonds $J_{ij}$. This means that multiple stable states do appear in the model of Axelrod and Bennett, and their references to the ideas of landscape, spin glass and frustration are justified. The problem how to find an optimal division of the set of countries is analogous to the problem how to find a local stationary state in the spin glass. What appears specific for the recent formulations of the Heider problem \cite{zth} is the possibility of a modification of the propensities as to get the balanced state.

\section{Summary}

To recapitulate, the dynamics of the Heider problem can be considered in different ways. Assuming that the character of ties does not change in time, one can ask about an optimal division. This question can be solved having the whole or only local information, with respect to the global balance or locally, allowing for a temporal increase of imbalance or not. Analogous set of problems is when the ties can be modified; then, the balanced state is expected as an output of the dynamics.

The list of approaches, provided above, can be treated as a set of exercises for students, who would like to work in the interdisciplinary area. If this attitude is more widespread, some methods and results can enter to the common knowledge as a useful starting point. Example giving, let us mention the case of a fully connected graph and all links negative as the initial state. Once we expect that the Heider balance will be attained, it follows from the symmetry of the problem that the appearing groups will be more or less of the same size. What seemed to be an artificial result in Ref. \cite{zth}, seems natural two years later \cite{red1}. Another example is the application of the genetic algorithm, mentioned in Ref. \cite{fre}. Once it is obvious that there is more than one solution of the problem, its formulation in terms of optimization methods is not a surprise. This is an argument to teach students of sociology not only programming, but also methods of artificial intelligence.

Having passed this short tour, we can ask if there is something new in the physicist's view on the Heider balance? At a first glance, the answer is no. The list of mathematical methods reviewed by Freeman \cite{fre} contains already an application of the genetic algorithm, and first application of the agent-based modelling to a sociological problem was not done by a physicist (a candidate could be Th. C. Schelling, Ref. \cite{who}). One could imagine, that the ironic statement of Johann Wolfgang von Goethe on mathematicians \cite{goe} applies well to what the sociophysics brings to sociology. However, extracting irony out from this sentence, we are left with the message with a strong optimistic charge. The optimism is in the possibility of reformulation of a social theory in the new language; in this case it is the language of statistical physics. And we know that new language constructs new reality. It seems that what we do in sociophysics is just preparing the way for new and maybe more succesful methods. 

{\bf Acknowledgements.}  Thanks are due to Dietrich Stauffer for his positive stimulation and useful reference data. 

\end{document}